\documentstyle[aps]{revtex}
\begin{document}
\twocolumn[\hsize\textwidth\columnwidth\hsize\csname
@twocolumnfalse\endcsname

    \title{\Large\bf Partial observables}
    \author{Carlo Rovelli\\ 
    {\small\it Centre de Physique 
    Th\'eorique de  
    Luminy, F-13288 Marseille, EU}\\
     {\small\it Department of Physics, Pittsburgh University, Pittsburgh  
     PA-15250, USA}}
     \date{\small\today}  \maketitle

     \begin{abstract} 
I discuss the distinction between the notion of {\em partial\/}
observable and the notion of {\em complete\/} observable.  Mixing up
the two is frequently a source of confusion.  The distinction bears on
 several issues related to observability, such as (i) whether time is
an observable in quantum mechanics, (ii) what are the observables in
general relativity, (iii) whether physical observables should or
should not commute with the Wheeler-DeWitt operator in quantum
gravity.  I argue that the {\em extended\/} configuration space has a
direct physical interpretation, as the space of the partial
observables.  This space plays a central role in the structure of
classical and quantum mechanics and the clarification of its physical
meaning sheds light on this structure, particularly in the context of
general covariant physics.

 \end{abstract} \vskip1cm]

\section{Introduction}

The notion of ``observable quantity", or ``observable", plays a
central role in many areas of physics.  Roughly, observable quantities
are the quantities involved in physical measurements.  They give
us information on the state of a physical system and may be predicted
by the theory.  In quantum mechanics, observables are represented by
self-adjoint operators.  In gauge theory, we make the distinction
between gauge-invariant quantities, which correspond to observables,
and gauge-dependent quantities, which do not not.

The notion of observable, however, raises a certain number of issues,
which have generated discussions in the literature.  In particular:
(i) Several papers discuss whether time is an observable in quantum
theory.  If time were an observable, it should be represented by a
self-adjoint operator $T$.  The spectrum of $T$ should be the real
line.  A well known theorem \cite{pauli} demands then its conjugate
variable, which is the energy, to have unbounded spectrum.  But energy
is bounded from below.  Therefore time cannot be an observable
\cite{timeop}.  But if time is not an observable, how can we measure
it?  (ii) There are several discussions on observability in general
relativity (see \cite{obsGR,observable} and references therein).  In
the literature one finds contradictory statements.  For instance, that
the metric tensor $g_{\mu\nu}(x)$ is not observable but a curvature
scalar $R(x)$ is observable; or that no local quantity such as $R(x)$
can be observable.  (iii) Observability is a source of lively debates
in quantum gravity \cite{observable,timeinqg}.  Observables must be
gauge invariant, therefore commute with the constraints, therefore, in
particular, with the Wheeler-DeWitt operator, and therefore they have
to be constant in the coordinate time $t$.  Thus, no quantity that
changes with $t$ can be observable.  This conclusion is considered
unreasonable by some \cite{Karel}.  Others (including myself)
\cite{observable,hypo,model,merced} argue that the observables in
quantum gravity are relative quantities expressing correlations
between dynamical variables.  But how can a correlation between two
non-observable quantities be observable?

I believe that in many debates of this kind there is a confusion
between two distinct notions of observability.  Mixing up these two
notions generates misunderstanding and conceptual mistakes.  In this
note, I try to clear up the source of this confusion.

The difference between the two notions of observability has to do with
localization in time and in space.  In a non-relativistic context, the
spacetime structure of the world is assumed to be fixed and simple. 
Because of this, the distinction between the two notions of
observability can be disregarded.  More precisely, the distinction is
replaced by the introduction of a fixed structure on the space of the
observables, and then it is safely ignored.  The fixed structure of
the space of the observables reproduces the fixed structure of
spacetime, as we shall see.  In a generally relativistic context, on
the other hand, the spacetime structure of the world is more complex,
and we cannot trade the distinction between different notions of
observability for a pre-established structure on the space of the
observables.  In such a context, ignoring the distinction between
different meaning of ``observable" leads to serious confusion.

Partial and complete observables are defined in Section \ref{PC}.  The
two notions are shown to be distinct and examples of the two are
given.  I then discuss the relevance of the distinction for different
contexts: general relativity (in Section \ref{GR}), quantum mechanics
(in Section \ref{QM}), and quantum gravity (in Section \ref{QG}).

The space of the partial observables is the {\em extended\/}
configuration space.  This space, and its associated extended phase
space, on which the hamiltonian constraint is defined, are often
presented as devoid of direct physical significance.  Instead, I argue
in Section \ref{ECS} that the extended configuration space has indeed
a direct physical interpretation: it is the space of the partial
observables.  This space plays a central role in the general structure
of mechanics, both at the classical and at the quantum level.  I
illustrate this role and argue that it provides a unifying perspective
that sheds light on the structure of mechanics, especially of general
covariant mechanics.

The distinction between partial and complete observables was discussed
in \cite{Boston}.  The distinction is sometimes implicitly used, but I
am not aware of any other explicit discussion on it in the literature.

\section{Partial observables and complete observables}\label{PC}

Let us start from the following two definitions\footnote{The
operational tone of the first definition does not imply any adherence
to operationalism here \cite{operat}: the reference to measuring
procedures is just instrumental for clarifying a distinction.}:
\begin{description}
    \item[{\em Partial observable:}] a physical quantity to which we
    can associate a (measuring) procedure leading to a number. 
    \item[{\em Complete observable:}] a quantity whose value can be
    predicted by the theory (in classical theory); or whose
    probability distribution can be predicted by the theory (in
    quantum theory).
\end{description}

At first sight, the two definitions might seem equivalent, but they
are not.  To see this, consider the following example.  Imagine we
have a bunch of cards in a box.  Each card has an upper and a lower
side (say, of different colors).  On each side, there is a number. 
Denote the upper number as $n$ and the lower number as $N$.  We
extract a certain number of cards from the box and we realize that
there is law connecting the two numbers: say $N$ is always a certain
function of $n$.  That is $N=N(n)$.  The law $N=N(n)$ gives us a
predictive theory for some observable quantities.  What are the
observables in this context?  Clearly, both $n$ and $N$ are {\em
partial observables}, according to the definition given above. 
However, neither of them is predictable, because at each new card we
extract we do not know which particular value of $n$, or which
particular value of $N$, will be found.  What is predictable is the
value of $N$ on the back of a card marked with a certain $n$. 
Therefore we have is one ``complete observable" $N(n)$ for each value
of $n$.  The ``complete observables" are $N(1),N(2),N(3),\ldots$

The example may seem artificial and unrelated to the structure of 
realistic physical theories, but it is not.  Indeed, realistic
physical theories have a structure similar to the one of the example:
the role of the ``independent" partial observable $n$ is played by the
quantities giving the temporal localization or the spatio-temporal
localization.  Consider for instance a very simple physical system, a
pendulum.  Assume oscillations are small and described by the equation
\begin{equation}
     \frac{d^2 q(t)}{dt^2}=-\omega^2 q(t). 
    \label{eq:HH}
\end{equation}
Now suppose we are in a (very simple) laboratory, and we want to check
the correctness of (\ref{eq:HH}).  What do we need?  Clearly we need
\emph{two} measuring instruments: one that gives us the pendulum
position $q$ and one that gives us the time $t$.  The theory cannot
predict the value of $t$.  Nor can it predict the value of $q$,
unless we specify that the value of $q$ we are interested in is the
one at a certain given time $t$.  Therefore, there are two partial
observables playing a role here: $q$ and $t$.  And there is one family
of complete observables: the observables $q(t)$, for any real value of
$t$.  It is sufficient to know the actual value of a few of these
complete observables (for instance $q(0)$ and $dq(t)/dt|_{t=0}$), in
order to be able to predict the value of all the others.

This may seem a rather pedantic account of observability in the
context of a non relativistic system.  Indeed, one usually says that
``$q$ is observable", leaving implicit ``yes, of course, one has to
say at which time the observation is made".  But, as mentioned, such
carelessness in defining observability is then paid for at a high
price in a generally relativistic context, where things are not simply
evolving in a fixed external time $t$ which can be measured by an
external clock, as in non-relativistic physics.  Let us therefore here
clearly distinguish between (i) $t$ and $q$ (without specified time),
which are partial observables, because there are measuring procedures
specified for them, but they cannot be predicted, and (ii) the family
of complete observables $q(t)$, which can be predicted.

Observe that the predictions of a mechanical theory can always be
expressed as {\em relations between partial observables}.  These
relations depend on a certain number of parameters, which label the
different possible histories of the system.  For instance, in the case
above the predictions of the theory are given by the following
relation between $t$ and $q$:
\begin{equation}
    f(q,t;A,\phi)=q-A \sin(\omega t +\phi) =  0. 
    \label{eq:solution}
\end{equation}
From this perspective, the partial observables $q$ and $t$ can be
taken as being on the same footing.  That is, they can be treated
symmetrically in the theory.  Observe, however, that in the example
considered the two partial observables $q$ and $t$ are not entirely on
the same footing.  The predictions of the theory can certainly be
expressed as a relation between the two, but this relation can be
solved for $q$ as a function of $t$, not for $t$ as a function of $q$. 
Accordingly, we call $t$ an {\em independent\/} partial observable and
$q$ a {\em dependent\/} partial observable.  As we shall see, in a
generally relativistic context such distinction between dependent and
independent partial observables is lost.

In a non relativistic system with $m$ degrees of freedom $q^{i}$, with
$i=1,\ldots,m$, there are in general $n=m+1$ partial observables:
$q^{a}=(t,q^i)$ with $a=1,\ldots,n$.  The space of these form the {\em
extended configuration space\/} of the system, which we denote $\cal
C$.  The predictions of classical mechanics can always be given as
relations between the extended configuration space variables, as in
(\ref{eq:solution}).  These relations depend on a certain number of
parameters $\alpha^j$ ($A$ and $\phi$ in (\ref{eq:solution})), which
label the different possible histories of the system.
\begin{equation}
    f(q^a;\alpha^j)=0.
    \label{eq:solutions}
\end{equation}
Classical mechanics and quantum mechanics can be formulated in a very
general and very clean form over the extended configuration space
$\cal C$.  Examples of such formulations are the Hamilton-Jacobi
formalism, the extended phase space formalism, the path integral
formalism and the propagator formalism \cite{RR}.  These formulations
stress the centrality of the notion of partial observable and show
that mechanics treats all partial observables on the same ground.  In
Section \ref{ECS}, we shall discuss some of these formulations and
their relation to partial observability.

Finally, consider a field theory, such as Maxwell electrodynamics.  A
dynamical variable is represented for instance by the Electric field
$E(\vec x,t)$.  The electric field at a given spacetime point $(\vec
x,t)$ can be predicted, and therefore it is a complete observable of
the theory.  In order to measure $E(\vec x,t)$, we need \emph{five}
partial observables.  Indeed, we may imagine that we have at our
disposal five measuring devices: a clock measuring $t$, an electric
field detector that measures $E$, and three distance measuring
devices, giving the three components of $\vec x$.  The complete
observable $E(\vec x,t)$ is composed by these five partial
observables.

\section{General Relativity}\label{GR}

Let us now move to a generally relativistic context.  For
concreteness, let us consider general relativity coupled with $N$
small bodies.  For instance, these bodies may represent the planets
and the satellites in a generally relativistic model of the solar
system.  The lagrangian variables can be taken to be the metric
$g_{\mu\nu}(\vec x, t)$ and, say, the bodies' trajectories
$X^\mu_{(n)}(\tau_{n})$, with $n=1,\ldots,N$ and orientations
$E^{a\mu}_{(n)}(\tau_{n})$ (a local tetrad on the $n-$th body,
$a=0,\ldots, 3$).  As well known, the meaning of the coordinates
$(\vec x, t)$ in general relativity is very different from their
peaning in pre-general-relativistic (pre-GR)
physics.\footnote{Einstein has described his 1912 to 1915 final
struggle for general relativity as a dramatic effort to understand the
new ``meaning of the coordinates".} Indeed, the coordinates $(\vec x,
t)$ do not represent observable quantities at all.\footnote{Unless one
fixes a physically interpreted gauge, in which case the discussion of
observability is a bit different, but the final conclusions are
unchanged.  See below.} That is, the general relativistic coordinates
$\vec x$ and $t$ are neither partial observables nor complete
observables.

The distinction between partial and complete observables, however, is
still present.  Consider some typical predictions of the theory.  For
instance, a prediction of the theory may be the following: tomorrow
morning, when the Sun is 5 degrees over the horizon, Venus will be
visible at 12 degrees over the horizon.  This is a well defined
prediction, and should thus refer to a complete observable.  The
complete observable is the angle $\alpha_{V}$ that Venus makes with
the horizon, at the moment in which the angle $\alpha_{S}$ of the Sun
with the horizon is 5 degrees.  Clearly, to verify this prediction we
need measuring procedures giving us the two angles.  Therefore the two
angles are partial observables.  The complete observable is the value
of $\alpha_{V}(\alpha_{S})$ for $\alpha_{S}=5^o$.

As a second example, we could replace $\alpha_{S}$ with the proper
time $\tau$ measured on Earth by a clock that started ($\tau=0$) at a
certain specified event $O$ (say, a certain eclipse).  Then again,
the proper time $\tau$ elapsed from the eclipse, or, equivalently,
the length $\tau$ of the Earth's worldline since the eclipses, is a
partial observable because it can be measured; but it is not a
complete observable, because it cannot be predicted.  Indeed, it is an
observable quantity used for localizing a spacetime point.

The key difference between general relativity and pre-GR physics as
far as observability is concerned is well illustrated by a third
example.  Consider the following (realistic!)  experiment.  A very
accurate clock is mounted on a satellite.  Say a satellite in the GPS
system.  The satellite broadcasts its local time and the signal is
received by the launching base, and compared with the time of an
equally accurate clock kept at the base.  As well known, the
discrepancy between the two due to generally relativistic effects is
easily observable using current technology.  Let $\tau_{s}$ and
$\tau_{b}$ be the signal received from the satellite and the local
clock reading.  General relativity can be used to predict the relation
between the two
\begin{equation}
    f(\tau_{s},\tau_{b})=0, 
    \label{eq:satellites}
\end{equation}
(once all the relevant initial data are known).  Again, we are in a
situation of two partial observables forming a complete observable. 
Now: which one of the two is the {\em independent\/} one?  In general,
(\ref{eq:satellites}) may not be solvable for either variable.  One
could say that $\tau_{b}$ has to be viewed as the ``natural"
independent variable, since this is ``our" time.  But one can equally
well say that the $\tau_{s}$ is the ``natural" independent variable,
since it provides an accepted standard of time \cite{GPS}\ldots \ \
Clearly we are in a very different situation from one with the two
partial observables $q$ and $t$ of the previous section.  There, we
had a clear distinction between an independent observable ($t$) and a
dependent one ($q$).  Here, $\tau_{s}$ and $\tau_{b}$ are truly on the
same footing.

The key difference between general relativistic physics and pre-GR
physics is the fact that in general relativistic physics the
distinction between {\em dependent\/} and {\em independent\/} partial
observables is lost.  A pre-GR theory is formulated in terms of
variables (such as $q$) evolving as functions of certain distinguished
variables (such as $t$).  General relativistic systems are formulated
in terms of variables (such as $\tau_{b}$, $\tau_{s}$, $\alpha_{V}$,
$\alpha_{S}$) that evolve with respect to each other.  General
relativity expresses relations between these, but in general we cannot
solve for one as function of the other.  Partial observables are
genuinely on the same footing.

What are the complete observables, in general, in this context?  A
complete observable is a quantity that can be predicted uniquely. 
Therefore it is a quantity which is well defined once we know the
solution of the equations of motion, up to all gauges (that is, which
not affected by the indetermination of the evolution).  Such a
quantity can be seen as a function on the space of the solutions
modulo all gauges.  This space is the physical phase space of the
theory $\Gamma$.  In the canonical formalism, $\Gamma$ can be obtained
as the space of the orbits generated by the constraints on the
constraint surface.  Any complete observable can thus be expressed as
a function on $\Gamma$.  Equivalently, it can be expressed as a
function on the extended phase having vanishing Poisson brackets with
all first class constraints, including, of course, the hamiltonian
constraint.  Vice versa, any function that commutes with all
constraints defines, in principle, a complete observable.

Partial observables are hard to construct formally in general, but it
is far easier to define and use them concretely.  For a recent
concrete construction of a complete set of partial and complete
observables in GR, see \cite{GPS}.

I close the section with a note on gauge-fixed formulations of GR. One
may fix the gauge by choosing coordinates that have a physical
interpretation.  More precisely, one may select a family of partial
observables (curvature scalars, scalar fields, dust variables, GPS
readings\ldots) and fix the coordinate gauge by tying the coordinate
system to these partial observables.  Within a formulation of this
kind, coordinates represent partial observables.  Furthermore, they
have a natural character of independent partial observables.  However,
this does imply that the independent partial observables are
determined by the theory, because the same physical situation can be
described by a different physical gauge choice, in which the role of
dependent and independent partial observables is interchanged.

\section{Quantum theory}\label{QM}

In quantum theory observables are represented by operators.  Which
observables are represented by operators: the partial or the complete
observables?  The answer is different in the Heisenberg picture
(evolving operators) and in the Schr\"odinger picture (evolving
states).  Let us start from the Heisenberg picture.  Here the
operators are time dependent.  For instance, in the quantum theory of
a harmonic oscillator in the Heisenberg picture, there is no position
operator $Q$, but only the operator $Q(t)$ that represents ``position
at time $t$".  This is immediately recognized as the operator
corresponding to the complete observable $q(t)$ discussed in Section
\ref{PC}.  In the Heisenberg picture operators are associated with
complete observables.

In the Schr\"odinger picture, there is an operator $Q$ associated with
the partial observable $q$.  However, specific predictions are not
given just in terms of this operator: we need the state as well, and,
in the Schr\"odinger picture, the state $\Psi(t)$ is time dependent. 
Thus, for instance, the expectation value $\overline
Q(t)=\langle\Psi(t) Q \Psi(t)\rangle$, which is a prediction of the
theory, is associated with the complete observable $q(t)$, as it
should, not with the partial observable $q$.  In order for the
Schr\"odinger picture to be meaningful, we need the theory to be
expressed in terms of a well defined independent partial observable
$t$: ``the external time".  In a theory such as general relativity,
where the dynamics expresses the relation between partial observables
that are on equal footing, the Schr\"odinger picture is not viable. 
More precisely, it will be viable only in special circumstances, in
which we can choose (arbitrarily) one of the partial observables as
the independent one and solve the dynamical relations expressing the
predictions of the theory in terms of this quantity.  In general, no
such quantity exists.  On the other hand, the Heisenberg picture
remains meaningful whatever the spacetime structure of the theory. 
Let us therefore return to the Heisenberg picture, which is far more
general.

In the Heisenberg context, consider the problem of whether there
should be a time operator in quantum theory.  The time $t$ is a
partial observable, not a complete observable.  Operators are
associated with complete observables, not with partial observables. 
Therefore it is against the tenets of quantum theory to search for an
operator corresponding to $t$.  Operators correspond to quantities
that are in principle predictable (such as $q(t)$), not to quantities
(such as $t$) that serve only to localize the measurement of a
predictable quantity in spacetime.  A quantity that is described by an
operator in quantum theory is a quantity such that there are states
that diagonalize it, namely such that there are physical situations in
which the outcome of a measurement of that quantity is certain: the
time $t$, on the contrary, can never be predicted.

In other words, quantum theory deals with the relation between $q$ and
$t$, and not with $q$ alone or $t$ alone.  Therefore it is meaningless
to search for the quantum theory of the $t$ variable alone.

Of course, the reading $T$ of a clock {\em can\/} be predicted, but
only {\em if\/} we first read {\em another\/} clock.  If we know that
the second clock indicates $t$, we can predict that the first clock
will read a certain $T$.  If we now take into account the fact that
the clock is a physical mechanical system and it is subject to quantum
fluctuations, then we can describe it in terms of an operator.  This
operator describes the complete observable $T(t)$.  There will be
quantum fluctuations described by generic states in the state space on
which this operator acts.  These fluctuations are not the quantum
fluctuations of one independent time variable.  They are the quantum
fluctuations in the observable correlation between {\em two\/} clock
variables.

\section{Quantum gravity}\label{QG}

In quantum gravity, operators corresponding to physical observables
must commute with the Wheeler-DeWitt constraint operator.  This
operator is the generator of evolution in the coordinate time $t$. 
Thus, physical observables must be invariant under evolution in $t$. 
This fact has raised much confusion.  How can observables invariant
under evolution in $t$ describe the evolution we observe?  The
question is ill posed, because it confuses evolution with respect to
the coordinate time $t$ and physical evolution.  In Section \ref{GR}
we have observed that in general relativity quantities like the proper
times $\tau_{b}$ and $\tau_{s}$ are partial observables and their
relative evolution is well defined.  Let us fix a value
$\tau_{b}=\tau$ of the first, and consider the corresponding value of
$\tau_{s}$.  (If there are several such values, take the highest). 
Call this value $T_{\tau}$.  That is $T_{\tau}$ is the highest number
for which
\begin{equation}
    f(T_{\tau},\tau)=0.  
    \label{eq:Ttau}
\end{equation}
where $f$ is the function in (\ref{eq:satellites}).  $T_{\tau}$ is a
complete observable.  It is the signal we receive from the satellite
when our local proper time at the base is $\tau$.  It describes the
change of the value of the received signal as the proper time at the
base passes.  This is a description of evolution.  At the same time,
this is a quantity independent of the coordinate $t$.  To see this,
recall that to calculate its value from a specific solution of the
Einstein equations, we first find the dependence of $\tau_{b}$ and
$\tau_{s}$ on the coordinate time $t$.  Namely we compute the
functions $\tau_{b}(t)$ and $\tau_{s}(t)$.  The form of these two
functions is gauge dependent: it changes if we use a different
coordinate representation of the same four geometry.  We then locally
invert the second function and insert $t(\tau_{s})$ in the first.  The
resulting $T_{\tau}\equiv\tau_{s}(\tau_{b}=\tau)$ is independent of
the coordinate $t$ chosen, and thus it is uniquely determined by the
equivalence class of solutions of the field equations under
diffeomorphisms.  It is a well defined on the space of these
equivalence classes, namely on $\Gamma$.  Equivalently, it can be
represented as a function on the extended phase space that commutes
with all the constraints, including the hamiltonian constraint.

Let us now come to a main objection that we want to address in this
paper, which is the following.  
\begin{description}
\item[{\em Objection:}] $T_{\tau}$ cannot be observable without
$\tau_{b}$ and $\tau_{s}$ being individually observable.  Thus
$\tau_{b}$ and $\tau_{s}$ are observable.  Observables must be
represented by physical operators.  $\tau_{b}$ and $\tau_{s}$ depend
on $t$ and do not commute with the hamiltonian constraint.  Therefore
in any quantum theory of gravity there should be physical operators
representing observables that do not commute with the Wheeler-DeWitt
constraint.
\end{description}
It should be clear at this point why this objection is wrong.  It
confuses partial and complete observables.  $\tau_{b}$ and $\tau_{s}$
are partial observables, and partial observables are not associated
with quantum operators in quantum theory (more precisely, in
Heisenberg picture quantum theory, which is the only one viable in
this context).

We close this section with an observation on the role of the
coordinates in the formalism of quantum gravity.  The general
relativistic spacetime coordinates $(\vec x, t)$ have no direct
physical interpretation.  In a gauge fixed context, they can be tied
to partial observables.  In any case, however, they do not represent
complete observables.  It follows that the idea that the coordinates
should be represented by quantum operators is not justified is the
light of quantum theory and general relativity alone.  Operators are
attached to complete observables, while spacetime coordinates are --at
best-- partial observables: they cannot be predicted, they serve only
to localize complete observables. 

Quantum theory deals with the relation between partial observables. 
It can deal with the relation between physical variables and
(gauge-fixed) coordinates $(\vec x, t)$.  But not with the value of
the coordinates alone.  Therefore it is meaningless to search for the
quantum theory of the $(\vec x, t)$ variables alone.

Non-commutative geometry approaches to quantum gravity search for a
mathematics capable of promoting the spacetime coordinates $(\vec x,
t)$ to a non-commuting operator algebra.  This approach is sometimes
motivated with the argument that quantum theory should require the
coordinates $(\vec x, t)$ to be represented by operators.  In the
light of the discussion above, I think that this motivation mistakes
partial observables and complete observables.  Non-commutative
approaches to quantum gravity are extremely interesting in my view,
both mathematically and the physically.  But I think that this
particular motivation is naive and not tenable.  Physical
non-commutativity of quantities related to physical localization and
geometry, on the other hand, is likely to follow from the fact such
quantities should, in fact, be functions of the gravitational field,
and therefore quantum dynamical variables.

\section{The extended configuration space}\label{ECS}

In this section I discuss the role of the partial observables in the
formal structure of mechanics.  I focus here on theories with a finite
number of degrees of freedom, leaving the extension to field theory to
the reader.\footnote{Examples of generally relativistic systems with a
finite number of degrees of freedom are provided for instance by
cosmological models, or by models as in \cite{model}.} This discussion
sheds light on the physical interpretation of certain structures, such
as the extended phase space of the fully constrained systems.  For a
more complete treatement of this subject, see \cite{foundation}; see
also \cite{merced}.  The central message of this section is double. 
First, that when the formalism is sufficiently general, partial
observables are the main quantities mechanics deals with.  Second,
that, in general, mechanics makes no distinction between dependent and
independent observables.  The distinction between independent and
dependent observables can be seen as an accident of the specific
dynamics of non-relativistic theories.  In the light of these two
observations, I think that the interpretation of general relativistic
theories becomes more transparent.

As observed in Section \ref{PC}, the partial observables of a
mechanical system form the extended configuration space $\cal C$. 
Recall that we denote the partial observables by $q^a$,
$a=1,\ldots,n$.  Dynamics can be given in terms of a first order
partial differential equation on $\cal C$, the Hamilton-Jacobi
equation
\begin{equation}
    C\left(q^a,\frac{\partial S(q^a)}{\partial q^a}\right)=0, 
    \label{eq:HJac}
\end{equation}
The function of 2$n$ variables $C(q^a,p_{a})$ determines the dynamics. 
One searches for an $n$-parameter family of solutions of this equation
$S(q^a,Q^a)$, where $Q^a$ are $n$ constants, and the predictions of
the theory are contained in equations (\ref{eq:solutions}), which
are obtained as follows
\begin{equation}
    f_a(q^a;Q^a,P_a)=\frac{\partial S(q^a,Q^a)}{\partial Q^a}
    -P_a=0. 
    \label{HJ}
\end{equation}
These form a $2n$-parameter family of $n$ relations between the
partial observables (not all independent).  The parameters $Q^a, P_a$
label the possible histories of the system: each history determines a
set of relations among partial observables.  These relations are the
physical predictions of the theory.  Notice that all partial
observables $q^a$ are treated on the same footing: the ``time" partial
observable, if present at all, is just a variable among the others. 
Notice also that the usual last step of the Hamilton-Jacobi
prescription, which is to invert (\ref{HJ}) for the dependent
variables, is not necessary from this point of view.

Let $\Gamma$ be the space of the histories.  Generically, there is one
history connecting any two points of $\cal C$.  Therefore $\Gamma$ has
dimension $2n-2$.  Since $Q^a$ and $P_a$ are $2n$ functions on
$\Gamma$, they over-coordinatize $\Gamma$ and there are 2 relations
among them.  Also, histories are one-dimensional, and therefore only
$n$-1 of the $n$ relations (\ref{HJ}) are independent.  The space
$\Gamma$ is the phase space of the system.  A point in $\Gamma$ is a
``state" of the theory, in the sense of ``Heisenberg state`"
\cite{Dirac}.  It represents a possible history of the system, not a
``state at a certain time".\footnote{Dirac argued repeatedly that the
Heisenberg notion of state is the good one, and the only one that
makes sense in a relativistic context.  See for instance Sec I.3 of
the first edition of \cite{Dirac}.  In later editions of this book
Dirac shifted the emphasis to the Schr\"odinger states, explaining (in
the Preface) that these, after all, are easier to work with in the non
relativistic context, although ``it seems a pity" to give the cleaner
notion.}

The function $f$ in (\ref{HJ}) is defined on the cartesian product of
the space of the partial observables with the space of the states.
\begin{equation}
    f: {\cal C}\times \Gamma \longmapsto I\!\! R^{n}. 
\end{equation}
The entire predictive content of a dynamical theory is in the surface
$f=0$ in the cartesian product of the space of the partial observables
and the space of the states.  For each point $q^a$ in $\cal C$, the
surface $f=0$ determines the set of states compatible with the value
$q^a$ of the partial observables.  For each state in $\Gamma$, the
surface determines a relation among the partial observables in $\cal
C$.

In the {\em special\/} case of a non-relativistic system, one of the
partial observables $q^a$ is the time $t$.  Let it be, say, $q^0$, and
call the other partial observables $q^i$ with $i=1,\ldots,m=n-1$.  In
this case the function $C(q^a,p_{a})$ has the {\em special\/} form
\begin{equation}
    C(q^a,p_{a})=p_{0} + H(q^i,p_{i}). 
\end{equation}
Therefore in this special case the Hamilton-Jacobi equation takes the
well known form 
\begin{equation}
    \frac{\partial S(q^i,t)}{\partial t} +H\left(q^i,\frac{\partial
    S(q^i,t)}{\partial q^i}\right)=0.
    \label{eq:HJt}
\end{equation}

The general Hamilton-Jacobi formalism has a nice geometrical
interpretation in the canonical framework.  Let us illustrate it, with
the purpose of discussing the meaning of the structures of generally
covariant hamiltonian systems.

Consider the cotangent bundle $T^{*}{\cal C}$ over the extended
configuration space, with canonical coordinates $(q^a,p_{a})$.  Call
it the extended phase space.  It carries the natural Poincar\'e
one-form $\theta=p_{a}dq^a$, and the symplectic form
$\omega=-d\theta$.  The dynamics is coded in a relation on $T^{*}{\cal
C}$:
\begin{equation}
    C(q^a,p_{a})=0,
    \label{eq:C}
\end{equation}
In the special case of a non-relativistic system, $q^a=(q^0,q^i)$ and
Equation (\ref{eq:C}) has the form
\begin{equation}
    C(q^a,p_{a})=p_{0}+H(q^i,p_{i})=0
    \label{eq:CH}
\end{equation} 
where $H$ is the Hamiltonian.  The variable $q^{0}=t$ is the time
variable, and its conjugate momentum $p_{0}=-E$ is (minus) the energy. 
The dynamics of the system is then coded in the relation (\ref{eq:CH})
which gives the energy as a function of the other coordinates and
momenta.

Equation (\ref{eq:C}) defines a surface
$\Sigma$ in $T^{*}{\cal C}$.  Call $\omega_{\Sigma}$ the restriction
of $\omega$ to this surface.  The ``presymplectic" two-form
$\omega_{\Sigma}$ is degenerate and has a null direction.  It is not
difficult to see that the integral curves of this null direction are
the solutions of the equations of motion of the system.\footnote{The
coordinate form of the relation $Y(\omega_{\Sigma})=0$ between
$\omega_{\Sigma}$ and its null vector field $Y$ is given by the
Hamilton equations.} The space of these curves is the physical phase
space of the system $\Gamma$ and carries a unique symplectic two-form
$\omega_{\Gamma}$ whose pull back to $\Sigma$ under the natural
projection $\pi:\Sigma\to\Gamma$ is $\omega_{\Sigma}$.  Let $P_{a}$
and $Q^a$ be coordinates that (over-)coordinatize\footnote{$2(n-1)$
coordinates are sufficient to coordinatize $\Gamma$.  For instance, one
can take initial coordinate and momenta at $t=t_{0}$.  We prefer to
use here $2n$ coordinates for reasons that will be clear below.  The
extra coordinates can be seen as the initial time $t=t_{0}$ and the
energy.  A change in the first amounts to a relabeling of the meaning
of the initial data.  The second is constrained by $C$.}
$\Gamma$ and define a one-form $\theta_{\Gamma}=P_{a}dQ^a$ such that
$d\theta_{\Gamma}=-\omega_{\Gamma}$.  Then (using the same notations
for the forms and their pull back), since on $\Sigma$ we have
$\omega_{\Gamma}=\omega_{\Sigma}$, it follows that
\begin{equation}
    d(p_{a}dq^a)=d(P_{a}dQ^a)
     \label{eq:d}
\end{equation}
or
\begin{equation}
    p_{a}dq^a=P_{a}dQ^a + dS
    \label{eq:S}
\end{equation}
where $S$ is a zero-form on $\Sigma$.  But let us pull the coordinates
$Q^a$ back onto $\Sigma$ and assume that the set $(q^a,Q^a)$ (over-)
coordinatizes $\Sigma$.  Then (\ref{eq:S}) gives
\begin{eqnarray}
\frac{\partial S(q^a,Q^a)}{\partial q^a} &=& p_{a},
    \label{eq:Sp}
\\ 
\frac{\partial S(q^a,Q^a)}{\partial Q^a} &=& P_{a}.
    \label{eq:Ss}
\end{eqnarray}
From (\ref{eq:Sp}) and (\ref{eq:C}) we obtain 
\begin{equation}
    C\left(q^a,\frac{\partial S(q^a,Q^i)}{\partial q^a}\right)=0, 
\end{equation}
which is the Hamilton-Jacobi equation (\ref{eq:HJac}) and can be used
to compute $S$; while (\ref{eq:Ss}) is the equation (\ref{HJ}) giving
the physical predictions from $S$.

What is the physical meaning of $S(q^a,Q^a)$?  Without loss of 
generality, we can choose the integration constants so that 
\begin{equation}
   S(q^a,Q^a=q^a)=0.  
\end{equation}
Fix a point $p$ on $\Sigma$, and consider the trajectory that starts
on $p$.  Along this trajectory $dQ^a=0$ and thus from (\ref{eq:S}) we
have $dS = p_a dq^a$.  Parametrize the trajectory with an arbitrary
time parameter $t$ and write $dS = p_a dq^a = p_a \dot{q}^a dt$.  The
canonical hamiltonian with respect to this parameter is null, and
therefore $p_{a}\dot{q}^a - L=0$. Therefore we have $dS = L dt $ along
each orbit $q^a(t)$ and
\begin{equation}
   S(q^a,Q^a)=\int_{Q^a}^{q^a} L(q^a(t)) dt.  
\label{SL}
\end{equation}
That is, $S(q^a,Q^a)$ is the action, computed over the physical
trajectory that joins the points with coordinates $Q^a$ and $q^a$.
In the case of a non-relativistic system, let $q^a=(q^i,t)$. Then 
$dS = p_a dq^a = p_i dq^i - H dt$.  Recall that $H =
p_{i}\dot{q}^i - L$, where $L$ is the Lagrangian.  Therefore $dS = p_i
dq^i - p_i dq^i + L dt = L dt $ along each orbit.  Thus,
\begin{equation}
   S(q^a,Q^a)=\int_{Q^i}^{q^i} L(q^i(t)) dt.  
\end{equation}
where the trajectory starts at time $t$ in $q^i$ and ends at time $T$
in $Q^i$.  That is, $S(q^a,Q^a)$ is still the action, computed over
the physical trajectory that joins the points with coordinates $Q^a$
and $q^a$.

Notice that from this point of view Hamilton's principal function and
Hamilton's characteristic function are identified.  More precisely,
$S(q^a,Q^a)$ is the principal function for the evolution in any
partial observable identified as the time $q^0=t$.  But it is also the
characteristic function of the evolution in an arbitrary parameter
time along the histories.  And it is also the principal function for
the evolution in such a time, since the hamiltonian that generates
this motion vanishes.  This compactification of the formalism is quite
remarkable.

In conclusion, the ingredients of mechanics can be taken to be solely
the extended configuration space $\cal C$ and the function $C$ on
$T^{*}{\cal C}$.  A mechanical system is determined by the pair
$({\cal C}, C)$.  The kinematics of a specific theory is determined by
the space of its partial observables $\cal C$; its dynamics is
determined by the constraint $C(q^a,p_{a})=0$ on the associated phase
space.  There is no need to single out a specific partial variable as
the time, nor to mention evolution.  Mechanics is a theory of
relations between partial observables.  No distinction between
dependent and independent partial observables is required.  This
distinction is an accident of non-relativistic theories, in which the
constraint $C(q^a,p_{a})$ happens to have the form (\ref{eq:CH}).

Why do I stress this fact?  Because generally relativistic theories
are formulated in terms of constraints such as (\ref{eq:C}) over an
extended configuration space.  It is sometimes claimed that the theory
can only be interpreted if one finds a way to ``deparametrize" the
theory, namely to select the independent variable, among the variables
$q^a$.  In the opposite camp, the statement is sometimes made that
only variables on the physical phase space $\Gamma$ have physical
interpretation, and no interpretation should be associated with the
variables of the extended configuration space $\cal C$.  Instead, I
have argued here that the variables of the extended configuration
space have a physical interpretation as partial observables.  In a
sense, they are the quantities with the most direct physical
interpretation in the theory.

Finally, consider quantum theory.  The Schr\"odinger equation, as well
as the Wheeler-DeWitt equation, are {\em both\/} partial differential
equations on the extended configuration space $\cal C$.  They can both
be obtained in general from the constraint (\ref{eq:C}), with no need
of distinguishing dependent from independent partial observables. 
Indeed, they are obtained as
\begin{equation}
C\left(q^a,-i\hbar\frac{\partial}{\partial q^a}\right)\psi(q^a)=0. 
    \label{eq:Sc}
\end{equation}
The usual physical scalar product on an appropriate space of the
solutions of this equation has an intrinsic meaning and does not need
the time variable to be singled out in order to be defined -- see for
instance \cite{Marolf,RR}.  All relevant physical predictions of the
theory can be extracted from the knowledge of the propagator
$W(q^a,Q^a)$, which satisfies
\begin{equation}
C\left(q^a,-i\hbar\frac{\partial}{\partial q^a}\right)W(q^a,Q^a)=0. 
    \label{eq:W}
\end{equation}
The propagator gives the probability amplitude for finding the
combination of partial observables $q^a$ if the combination of partial
observables $Q^a$ was previously observed.  Virtually all predictions
of quantum mechanics can be formulated in this covariant manner, on
the extended configuration space.  This is discussed in detail in
\cite{RR}.  As well known, in the limit of small $\hbar$ the
Schr\"odinger equation (or the Wheeler-DeWitt equation) goes over into
the Hamilton-Jacobi equation, and the propagator $W(q^a,Q^a)$ is given
to first order just by the exponential of the action $W(q^a,Q^a)\sim
\exp\{iS(q^a,Q^a)/\hbar\}$.

\section{Conclusions}

I have observed that the notion of observable is ambiguous, and I have
discussed the distinction between partial observables and complete
observables.  This distinction clarifies a certain number of issues
related to observability.  In particular, I have examined the role
played by this distinction in general relativity, in quantum mechanics
and in quantum gravity.

The partial observables form the extended configuration space $\cal
C$.  This space seems to be a natural home for classical and quantum
mechanics.  The two theories admit a clean formulation over this
space, which is sufficiently general to deal naturally with general
relativistic systems.  

A mechanical system is a pair $({\cal C},C)$.  The space of the
partial observables $\cal C$ describes the kinematics of the theory. 
$C$ is a function on $T^{*}{\cal C}$ that determines the dynamics. 
Classical dynamics is about relations between partial observables. 
These relations depend on a certain number of parameters, which label
the (time independent) states of the system.  The space of these
states is the phase space $\Gamma$.  The predictions of the theory are
therefore given by a surface $f=0$ on ${\cal C}\times\Gamma$.  The
surface $f=0$, as well as $\Gamma$, are determined by the pair $({\cal
C},C)$.  (The general structure of classical and quantum mechanics in
this language is discussed more in detail in \cite{foundation}.)

By fixing a subset of partial observables (one for a mechanical
system, four for a field theory), the other partial observables are
determined as functions on $\Gamma$.  This defines the complete
observables of the theory, whose value can be predicted uniquely if
the state is known.

Quantum mechanics gives the probability amplitude $W(q^a,Q^a)$ for 
measuring the combination of partial observables $q^a$ after having
measured the combination $Q^a$.  Alternatively, it gives the
probability distribution for the different possible outcomes of a
measurement of the complete observables.  These are represented by
self-adjoint operator over the (Heisenberg) state space.

No distinction between independent and dependent partial observable is
required.  The different partial observables can be viewed as being on
the same footing.  This formulation of mechanics does not require the
notion of external time.  It is therefore appropriate for general
relativistic systems, which are not formulated in terms of evolution
in time.

\centerline{------------------}

I thank Julian Barbour for extensive discussions and for help with the
language.  This work was partially supported by NSF Grant PHY-9900791.

\end{document}